# Learning Science and the Illusion of Understanding:

# Exploring the Effects of Integrating Learning Tasks after Explainer Videos


Madeleine Hörnlein[a] and Christoph Kulgemeyer[b]*,

[a] *Physics Education, Department of Physics, University of Paderborn, Germany,*
*https://orcid.org/0000-0002-4220-930X*

[b] *Institute of Science Education, Physics Education Group, University of Bremen, Germany,*
*https://orcid.org/0000-0001-6659-8170*

*[*]Corresponding author: Christoph Kulgemeyer, Institute of Science Education, Physics*
*Education Group, University of Bremen, Otto-Hahn-Allee 1, 28359 Bremen, Germany, E-*
*mail: kulgemeyer@physik.uni-bremen.de*


*Madeleine Hörnlein* is a PhD candidate in physics education at the University of Paderborn. Her research focuses on physics explainer videos.

*Christoph Kulgemeyer* is a professor of physics education at the University of Bremen. He works on physics teachers' professional knowledge, instructional explanations, and reflection skills. His latest work focuses on how to learn effectively using physics explainer videos.

**Funding information**: No funding was received for conducting this study.



# Learning Science and the Illusion of Understanding:

# Exploring the Effects of Integrating Learning Tasks after Explainer Videos


Explainer videos are increasingly used to support science learning. While prior research has demonstrated their potential, studies have also identified limitations—particularly their tendency to foster an *illusion of understanding*, where learners overestimate their grasp of a topic despite gaps in their actual knowledge. Pairing explainer videos with cognitively engaging elements may help mitigate this effect. This paper reports two experimental studies examining the immediate and long-term effects of learning tasks following a physics explainer video on learners' illusion of understanding. Study 1 ($N = 244$ physics students) compared high-level learning tasks with watching the video alone. Study 2 ($N = 175$) compared high-level and low-level tasks. Results show that high-level learning tasks significantly reduce the illusion of understanding immediately after the intervention compared to watching the video alone ($t(88) = 6.50$, $p < .001$, $d = 0.69$). Over the long term, both high- and low-level tasks are similarly effective. Learners with lower prior content knowledge are more susceptible to an illusion of understanding. We conclude that explainer videos should not be used in isolation in science classrooms. To prevent misjudged understanding—particularly among students with limited prior knowledge—they should be combined with cognitively demanding follow-up activities.

**Introduction**

Explainer videos (sometimes also referred to as instructional videos) are short video presentations designed to convey explanations of specific topics or principles. Such videos are widely available on online platforms and are frequently used both for self-directed learning and in formal educational settings (MPFS, 2021). In science education, teachers increasingly incorporate explainer videos into their instruction, for example, in approaches like the flipped classroom. Given their growing role in science teaching, understanding how to use them effectively in learning environments is crucial.

Numerous studies have examined the potential of explainer videos to support learning (e.g., Lloyd & Robertson, 2012; Brame, 2016). However, research has also highlighted their limitations—most notably, their tendency to induce a so-called *illusion of understanding* (Paik & Schraw, 2013). This phenomenon arises when learners believe they have grasped a topic, while their actual understanding remains superficial (Prinz, Golke & Wittwer, 2019). Such illusions can negatively affect science learning in multiple ways. When students are under the impression that they have already mastered a concept, they may perceive further instruction as redundant, which can reduce their level of cognitive activation. Since cognitive activation is a key component of effective teaching, this may directly impair instructional quality (Kunter et al., 2013). Moreover, self-directed learning contexts—where explainer videos are commonly used—require learners to accurately assess their own understanding. If learners overestimate their comprehension, they may prematurely terminate the learning process (Kulgemeyer, Hörnlein & Sterzing, 2022). In essence, an illusion of understanding is detrimental to learning outcomes because it involves a false belief that one has understood a concept, leading to an overestimation of one's knowledge. Such beliefs are known to guide behavior and are often resistant to change (Fives & Buehl, 2012)—in this case, influencing how learners respond to further instruction. Overestimating one's knowledge can reduce the



perceived relevance of additional learning opportunities and diminish cognitive activation (Kulgemeyer, Hörnlein & Sterzing, 2022). Furthermore, systematic overestimation of understanding makes conceptual change highly unlikely, as dissatisfaction with one's current knowledge is a crucial prerequisite for conceptual change (Posner et al., 1982; Duit & Treagust, 2012).

Previous research suggests that explainer videos should not be used as standalone learning tools due to the high risk of fostering an illusion of understanding (Wittwer & Renkl, 2008; Kulgemeyer, 2018). One potential approach to mitigating the illusion of understanding is the use of *learning tasks*. From a constructivist perspective, engaging with learning tasks encourages deeper processing of the explained content, potentially leading to more effective learning (Webb et al., 2006; Kulgemeyer & Geelan, 2024). However, it remains unclear whether learning tasks are also an effective means of reducing illusions of understanding.

This article presents two related experimental studies investigating the role of learning tasks following an explainer video in reducing the illusion of understanding and fostering conceptual knowledge acquisition in physics. Two types of learning tasks were examined: (1) low-level tasks that focus on retention of the video content, and (2) high-level tasks that require deeper cognitive engagement by applying the video content to novel situations. Both studies employed a pre-test, post-test, and follow-up test design to assess both immediate and long-term effects. The first study compares the impact of watching an explainer video alone versus combining it with a high-level learning task on the development of an illusion of understanding. The second study contrasts the effects of two types of learning tasks (high-level versus low-level) on the development of an illusion of understanding (cf. Section "Research Questions and Hypotheses"). By systematically examining the role of learning tasks in mitigating the illusion of understanding, these studies aim to provide evidence-based recommendations for the effective use of explainer videos in science education.



**Literature Review**

*Learning with explainer videos*

An explainer video is a short film that has usually a length of 6-9 minutes (Guo et al., 2014). Such videos can be found on participatory video platforms and are used for the purpose of teaching and learning. According to Wolf (2015), explainer videos (also referred to as explanatory videos or instructional videos) are self-produced films that explain how to do something or how something works or in which abstract concepts are explained.

Based on existing research on instructional explanations in science teaching (Wittwer & Renkl, 2008), criteria for particularly effective explainer videos were derived and tested in previous research (e.g., Kulgemeyer, 2018). Brame (2016) establishes criteria based on cognitive load theory (e.g., Sweller, 1994) and multimedia learning (Mayer, 2001). She recommends, among other things, that elements that promote cognitive activation can positively influence learning with videos (Brame, 2016). A number of factors have been found to enhance the quality of explanations including clear structure, adaptation to the learner's needs, tools for adaptation, concise explanations, emphasis on relevance, follow-up learning tasks, and the introduction of new, complex principles (Kulgemeyer, 2018).

Learning with explanations in general and explainer videos in particular is just effective when they are included into a constructivist learning environment (Wittwer & Renkl, 2008; Kulgemeyer & Geelan, 2024). In addition to the possibilities of teachers and learners producing videos themselves, there are various methods of integrating pre-existing videos from online platforms into school-based learning, such as blended learning or the flipped classroom. These approaches have gained prominence in recent years. In blended learning, an explainer video is viewed while problem-solving or actions are performed. The flipped classroom concept involves instruction through an explainer video at home, reversing conventional classroom instruction (Van Alten, Philiex, Janssen & Kester, 2019). In-person



class time is then utilized to apply the learned material with the teacher's guidance. This approach allows for a more efficient use of class time and promotes active learning (Bishop & Verleger, 2013). However, this only works if not only the learning location has changed but the entire teaching concept is adapted accordingly.

While learning with (online) explainer videos offers potential, it also has limitations. For instance, online platforms may prioritize click figures over optimal content presentation. Additionally, studies have demonstrated that the platform's inherent markers, such as click figures or likes, do not align with the objective quality of the explainer video according to a grid of objective criteria (Kulgemeyer & Wittwer, 2023; Kulgemeyer & Peters, 2016).

However, even good explainer videos according to the aforementioned lists of effectiveness criteria do not automatically guarantee learning success. Rather they merely increase the probability that something will be understood (Kulgemeyer & Geelan, 2024). This is because the learner must construct the meaning from the videos themselves and make connections. Adapting the explanation to the learner's knowledge level is, therefore, essential (Wittwer & Renkl, 2008).

Considering this background, it can be posited that learning with explainer videos cannot stand alone but must be integrated into the active construction of knowledge, and the knowledge must be applied in order to sustainably learn from it (Wittwer & Renkl, 2008). Another important limitation of explainer videos, particularly in science, is their potential to create an illusion of understanding.

### *Learning science and the 'illusion of understanding'*

It has often been emphasized that learners should not be left to their own devices with an explanation, as there is a risk that the explanation may create a so-called illusion of understanding (Wittwer & Renkl, 2008). Even good explanations or explainer videos carry the risk of an illusion of understanding (Prinz, Golke & Wittwer, 2018). "Illusion of



understanding" refers to the phenomenon that, after receiving an explanation, the accuracy of learners' belief of understanding is often low (Wittwer & Renkl, 2008). However, engaging in tasks after the explanation has been described as a promising way to counterbalance this effect (Chi, de Leeuw, Chiu, & LaVancher, 1994). Kulgemeyer, Hörnlein and Sterzing (2022) defined this concept as follows: 'An illusion of understanding is the mistaken belief that (1) an instructional explanation is (a) a scientifically correct and (b) a high-quality explanation in terms of comprehensibility, and (2) the learners themselves (c) understood the concept, and (d) do not require further instruction.' (Kulgemeyer, Hörnlein & Sterzing, 2022, p. 4) The learner's assessment of their own understanding can initially be described neutrally as the belief of understanding. An illusion of understanding arises only when this belief diverges from one's actual understanding. For example, previous research has shown that presenting common misconceptions as scientifically correct explanations can lead to a high perceived level of understanding—although such confidence is not warranted (Kulgemeyer & Wittwer, 2023). This understanding builds on the idea that beliefs—in contrast to attitudes— can be confirmed or rejected on the basis of objective criteria (Eagly & Chaiken, 1993), in this case by an objective assessment of knowledge.

Research on metacognition and confidence in one's own performance has a rich tradition in psychology and mathematics education (Schneider & Artelt, 2010). Metacognition also includes "monitoring and self-regulation of one's own cognitive abilities" (Linge, Lenhart, & Schneider, 2019, p. 587). As such, it also encompasses monitoring one's conceptual understanding of explanations, which makes the illusion of understanding, as defined above, a metacognitive belief. Research in mathematics education has often focused on monitoring during problem-solving activities (Linge, Lenhart, & Schneider, 2019), with monitoring defined as the continuous observation of one's own cognition in order to regulate it (Schoenfeld, 1985). When monitoring fails, being overly (or insufficiently) confident may



lead to erroneous regulation of cognitive processes. This is also the case for an illusion of understanding, which can hinder deeper engagement with the content of an explanation (Kulgemeyer & Wittwer, 2018). Typically, monitoring is assessed by comparing an individual's predicted performance on a task with their actual performance, although various approaches have been discussed (Linge, Lenhart, & Schneider, 2019). An illusion of understanding cannot be assessed through predictive performance: Predicting performance requires a known—or at least accessible—task (Linge, Lenhart, & Schneider, 2019). By definition, an illusion of understanding goes beyond *predicting* performance on a specific problem: it includes evaluating one's own understanding of an explanation or instructional situation more broadly (criteria c and d), as well as evaluating the quality of the perceived explanation itself (criteria a and b). Therefore, it can only be assessed after the instructional situation—in the present study, after viewing an explanation video.

The illusion of understanding is an established concept in the context of instructional explanations (Wittwer & Renkl, 2008) and adds an important perspective to research on metacognition. In a nutshell, an illusion of understanding is not merely a judgment of understanding but is closely tied to how the explanation is perceived. Instead of representing a general self-evaluation of knowledge, it combines the perception of the instructional situation with the perception of its consequences for one's own learning progression. Only when the explanation is perceived as correct, the belief that one has understood the topic is high, and at least one of these two assumptions is objectively wrong, can we speak of an illusion of understanding. For example, a learner's metacognitive judgment of having understood a topic may be very high; however, if the explanation itself contains incorrect content, this still constitutes an illusion of understanding (as in Kulgemeyer & Wittwer (2018))—even though the judgment of having understood the incorrect explanation is, in a narrow sense, accurate.



Illusions of understanding refer to an observable phenomenon that arises in the context of explanations. It is not yet clear how the psychological mechanisms underlying this phenomenon operate, but the interplay of three distinct mechanisms may help to explain its occurrence: the Dunning–Kruger effect, the illusion of explanatory depth, and—particularly in the context of explainer videos—the shallowing hypothesis.

The Dunning-Kruger effect (Dunning et al., 2003) describes the tendency for learners to overestimate their abilities (see also Bol and Hacker (2012) for similar findings in mathematics). Those who perform poorly are especially challenged by the metacognitive task of evaluating their performance (e.g., Dunning, 2011). The overestimation of one's own understanding may also be associated with a phenomenon known as "illusion of explanatory depth" (Rozenblit & Keil, 2002). This notion posits that individuals who have comprehended a phenomenon at a specific level may erroneously conclude that they have achieved a comprehensive understanding of the phenomenon in its entirety. This illusion is particularly prevalent in the explanation of mechanical devices or natural processes, particularly when the explanation employs an abstract construal level (Alter, Oppenheimer & Zemla, 2010). This seems to be of a particular importance for explainer videos: explainer videos usually illustrate an abstract principle (the construal level) by using examples. However, the feeling of superficially understanding how the examples work can lead to an illusion of explanatory depth. Moreover, understanding how an underlying principle is connected to the presented examples should not be mistaken for having a comprehensive grasp of that principle. Research has shown that even a scientifically accurate video may prompt learners to overestimate their understanding. The "Shallowing hypothesis" (Salmerón, Sampietro & Delgado, 2020) offers a potential explanation. In accordance with this hypothesis, learners – at least K-12 learners – often consume social media and online videos during their leisure time with the intention of achieving immediate gratification (Salmerón, Sampietro &



Delgado, 2020). This may contribute to a superficial processing of explainer videos when they are utilized in the learning process.

Additional to the medium itself the media design of explanations has been examined for its connection with the occurrence of an illusion of understanding. For instance, the impact of motion pictures or animations on people should be considered. Such visual elements may prove distracting from the explained concept, as they tend to draw attention to the surface level of the concept being explained (Lowe, 2003). Additionally, they may hinder the ability to assess the effectiveness of the explanation and to judge one's own understanding and effort (Wiley, 2019). The effect that animation techniques create fun can create the illusion that learning is easy and that the presented content is learned quickly (Laaser & Toloza, 2017). Overall, they may influence the feeling of how much is actually understood (Salomon, 1984). Representational animations create an illusion of understanding, especially for learners with low general learning scores (Paik & Shraw, 2013). Nevertheless, a subsequent investigation indicates that the illusion of understanding is not solely contingent on the design of the media, but also on the comparative experimental observation of explanations in text form versus explanations in video form. This revealed no difference in learning success or in the belief of understanding (Kulgemeyer, Hörnlein & Sterzing, 2022). However, since only an introduction to a complex topic from physics was provided in each case, learners must still comprehend the entire topic, suggesting that an illusion of understanding may also be present.

It is reiterated that teaching cannot be limited to the mere presentation of an explanation; rather, additional learning tasks could be employed to demonstrate to learners the gaps in their understanding (Webb et al., 2006; Kulgemeyer, Hörnlein & Sterzing, 2022). Research indicates that learners may benefit from explaining the recently acquired content, thereby facilitating the identification of gaps in their understanding (Fernbach et al., 2013).



To elucidate for learners what they had not yet comprehended, learning tasks could be an efficacious approach to countering the illusion of understanding.

### Learning tasks

A learning task or instructional task is an interface between the learner and the content to be learned. They serve to activate or control learning processes (Richter, 2012). In general, a task can be defined as verbal or written instruction to perform a specific action. In physics lessons, tasks refer to a factual activity with the aim of reproducing, applying, or transferring physics knowledge (Kauertz & Fischer, 2021). It is important to distinguish between learning tasks that have the function of initiating and controlling learning processes and performance tasks that are intended to test knowledge (Kauertz & Fischer, 2021).

In addition, considerations such as cognitive structuring, linguistic simplicity, semantic redundancy, and brevity and conciseness can facilitate the comprehension of task texts. Laumann et al. (2024) note a lack of knowledge about how specific design elements of learning tasks, such as openness or contextual reference, affect learning. In general, there is relatively little consistent literature that sheds empirical light on quality criteria for effective learning tasks. Regarding their effectiveness, learning tasks are often categorized into low-level tasks and high-level tasks. This distinction is closely related to Bloom's taxonomy (Bloom, 1956). In this context, *low-level tasks* are sometimes also referred to as performance-oriented tasks and only require knowledge retrieval from memory. In contrast*, high-level or structure-oriented tasks* are all tasks that go beyond mere reproduction, e.g., the application of a principle (Brophy & Good, 1986; Renkl & Helmke, 1992). A higher level of activity and more cognitive steps are required for the latter. A German study indicates that low-level tasks are the predominant type set in physics lessons, with a slight overall variation in cognitive processes (Schabram, 2007). Studies have not identified a clear tendency towards the learning effectiveness of the two task types (Samson et al., 1987). This is also demonstrated by Renkl



and Helmke (1992) in relation to general learning success in mathematics. Nevertheless, a more recent study in biology education has demonstrated that instructional tasks requiring higher levels of cognitive engagement have a beneficial impact on learner's conceptual knowledge (Förtsch et al., 2018).

It may therefore be concluded that high-level tasks provide greater support for conceptual understanding or learning success. Regarding the illusion of understanding, we are not aware of research examining the impact of high-level versus low-level tasks. In this regard, low-level task stays close to the presented video (thus, demanding mere retention of the video content), while a high-level task introduces something new and requires more abstract thinking, which could counteract the illusion of explanatory depth (Alter et al., 2010). The potentially higher cognitive engagement resulting from the combination of videos and subsequent tasks, compared to merely watching videos, has been hypothesized to lead to a reduction in the illusion of understanding (Kulgemeyer, 2018).

## Research Questions and Hypotheses

Integrating learning tasks after watching explainer videos might lead to more realistic self-assessments and a reduction in the illusion of understanding. However, the impact of learning tasks on the illusion of understanding remains unclear, both in the short term and over the long term. This issue is particularly relevant for science education, as explainer videos are commonly used in instruction. While research has established criteria for designing effective videos, evidence-based practices for their effective integration into teaching are largely unknown.

To address this, we conducted two related experimental studies. The first study compares the effects of high-level learning tasks on the illusion of understanding with those of watching a video alone. Because low-level learning tasks are more common in physics



classrooms, we also conducted a second study that compares the effects of low-level and high-level learning tasks on the illusion of understanding, directly contrasting the two.

Thus, Study 1 investigates the following research question:

*RQ1: Does completing a high-level learning task after watching an explainer video reduce the illusion of understanding compared to watching the video?*

Based on the literature, we propose the following hypotheses:

H1a: Combining explainer videos with high-level learning tasks reduces the illusion of understanding compared to watching the video alone.

H1b: Learners with low knowledge experience a higher illusion of understanding than learners with high knowledge.

H1a is based on the idea that explanations are most effective when learners recognize gaps in their knowledge (Acuña, 2011; Fernbach et al., 2013). Kulgemeyer, Hörnlein, and Sterzing (2022) found that the shallowing hypothesis plays a role as a psychological mechanism underlying the illusion of understanding. In contrast, the present paper focuses on the Dunning–Kruger effect as a possible mechanism; accordingly, H1b investigates the Dunning–Kruger effect as a potential explanation for the occurrence of an illusion of understanding. While Study 1 focuses on whether high-level tasks reduce the illusion of understanding compared to no task, Study 2 examines whether low-level tasks have a similar or weaker effect. Thus, study 2 compares two types of learning tasks: low-level and high-level tasks. The theoretical considerations above suggest that high-level tasks are more effective in reducing the illusion of understanding; therefore, Study 1 focuses on them. However, if low-level tasks also contribute to reducing the illusion of understanding, this would have greater implications for physics instruction, as they are easier for teachers to implement spontaneously (since asking for retention of the video content is relatively



simple), whereas high-level tasks require more preparation. The research question for Study 2 is:

*RQ2: Do high-level or low-level tasks reduce the illusion of understanding more effectively than watching the video alone?*

The following hypotheses are derived from the literature:

H2a: High-level tasks lead to a greater reduction in the illusion of understanding than low-level tasks.

H2b: The reduction in the illusion of understanding over the long term persists more strongly for both types of learning tasks compared to watching just a video.

These hypotheses are grounded in theoretical considerations regarding the illusion of understanding. H2a is derived from the notion that low-level tasks remain closely aligned with the video content, whereas high-level tasks require deeper reflection, as they introduce novel or more abstract elements, thereby counteracting the illusion of explanatory depth (Rozenblit & Keil, 2002). H2b follows from the assumption that once the illusion of understanding is reduced, it is unlikely to re-emerge in the absence of further instruction beyond the explainer video and the learning task. Thus, the expected short-term reduction in the illusion of understanding (H2a) is hypothesized to persist over time.

Regarding control variables to ensure the comparability of the experimental and control groups, learners who are highly convinced that they are good at physics may also be easily convinced that they understand a physics topic well. Thus, because students' physics self-concept may influence their belief of understanding, we included it as a control variable.



## Methods

### *Context*

Both studies were conducted in an introductory physics course for prospective primary school teachers at a German university; however, they took place in different years and were therefore conducted independently. This course is designed for first-semester students and covers the fundamental principles of school physics. The course includes a lecture on the fundamentals of physics and a laboratory in which students conduct experiments related to the lecture topics. The course concludes with an assessment at the end of the semester. The study was conducted at the beginning of the semester, and there had been no prior instruction on the concept of energy. During the study, learning was individualized, and no in-class interactions were permitted.

Figure 1 illustrates the connection between study 1 and 2.

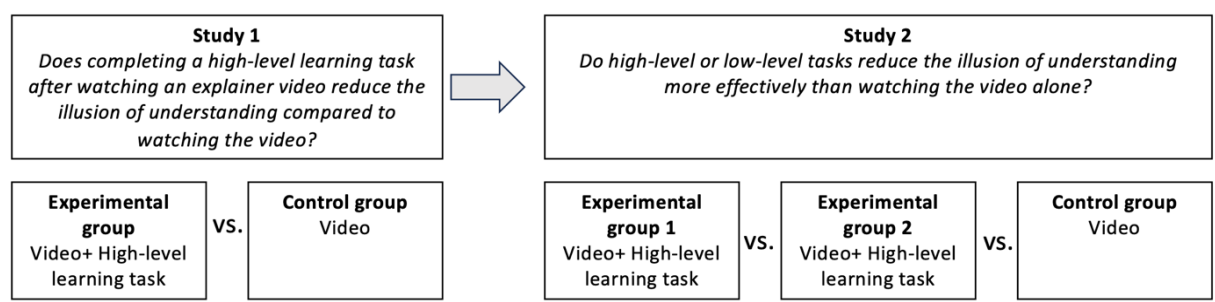

**Figure 1. Flowchart illustrating the connection between study 1 and 2.**

### *Participants*

In all phases of both studies, participation was voluntary. This led to variability in the number of complete datasets across the different phases of the study.

### *Study 1*

The sample consists of 244 students who participated in at least one of the three testing sessions. In the results section we will report the sample size for each analysis more in detail. The majority of participants (152) had not taken a physics course in the last three years



of school, which is the academic preparation phase of the German school system. This is consistent with the typical experience of students in this course. Of the participants, only 41 had studied physics at an advanced level in school, while 51 decided to not provide information on their prior physics education. Of the participants, 22 identified as male, 167 as female, one as non-binary, and 54 did not disclose their gender. This gender distribution is consistent with the demographic composition of the student population enrolled in preservice elementary teacher programs, which tends to overrepresent female students.

*Study 2*

In Study 2, 175 students participated. The sample for the second study was similarly composed to that of the first study in terms of demographics (female: 136; male: 25; non-binary: 1; no disclosure: 54) and prior education in physics (advanced level: 41).

**Materials: Explainer video and learning tasks**

For the use in both studies, we created an explainer video to introduce the concept of energy in a qualitative manner[1]. All participants across all groups watched the video during the intervention; therefore, we do not expect the video's characteristics to have influenced the findings. It was designed to resemble typical YouTube explainer videos, allowing us to investigate the effectiveness of this format in a controlled laboratory setting.

The video content aligns with Duit's Energy Quadriga and the Learning Progression of Energy (Neumann et al., 2012). In accordance with these frameworks, energy is introduced through four key properties that help learners understand its fundamental nature:

1. Energy can exist in different forms and can be transformed from one form into another.

2. Energy can be transferred from one location to another.

---

[1] Explainer Video in German Language: *blinded*



3. Some energy becomes less useful during transformations (energy degradation).

4. The total amount of energy in a system remains constant; energy is neither created nor destroyed (conservation of energy).

This conceptualization is consistent with curricula in Germany (e.g., KMK, 2020). The video has a duration of 7 minutes and 30 seconds, making it comparable to typical YouTube explainer videos (Wolf, 2015). It follows a validated framework for effective explainer videos (Kulgemeyer, 2018), presents several illustrative examples of the four main energy properties (e.g., jumping from a diving tower), and uses accessible language appropriate for the target audience—for instance, by directly explaining scientific terminology. The English translation of the video script is provided in the appendix.

Task:

The explainer video you have watched outlines the four main characteristics of energy.

a) Name the characteristics of energy mentioned in the video and explain them.

b) In the video, the characteristics of energy are introduced using specific examples. Explain the four main characteristics of energy again using the examples given in the video (bicycle with a dynamo and lamp).

c) At the end of the video, the four main characteristics of energy are applied to the example of a diver climbing a diving platform. Describe in your own words the main characteristics of energy using the specific example of the diver.

Task:

The explainer video you have watched outlines the four main characteristics of energy.

Now, apply these characteristics to the following examples:

1. A car is driving up a hill. The process involves friction.

2. Electrical energy from a wind turbine is used in a household to light a lamp.

**Figure 2. Overview of the learning tasks. The left panel shows the low-level task, the right panel depicts the high-level task. The tasks were originally developed in German language.**

To investigate the effect of learning tasks on the illusion of understanding when using explainer videos, two distinct learning tasks were developed (figure 2). One task was designed to replicate the video content (low-level task). The second task, which can be classified as a high-level task, requires the application of the explained principle to various contextual examples. The tasks vary in cognitive complexity and can be classified as either low-level or high-level tasks.



***Instruments/Measures***

Both studies evaluated the participants' conceptual knowledge about energy in the pre-test, post-test, and follow-up test. Additionally, it assessed their belief of understanding in post-test I (between video and task), post-test II (after task), and follow-up test.

Furthermore, the study included demographic information and control variables (gender and academic self-concept in physics), which were assessed during the pre-test.

*Control variables*

Gender was measured based on participants' self-identification. Academic self-concept in physics was assessed using a 5-point rating scale consisting of six items (study 1: $\alpha = 0.91$, study 2: $\alpha = 0.89$ e.g., ' I learn new material in physics quickly.'). The scale was adapted from Kulgemeyer and Wittwer (2023) and was also used in the work of Kulgemeyer, Hörnlein and Sterzing (2022). The scale is included in the appendix of the present paper.

*Illusion/Belief of understanding*

In order to operationalize the illusion of understanding, we employed a scale to measure the belief of understanding. Only in correlation with the actual knowledge, which we also measured, can inferences about a possible illusion of understanding be drawn from the belief of understanding scale. The instruments employed for the belief of understanding were consistent with those utilized by Kulgemeyer and Wittwer (20023) and Kulgemeyer, Hörnlein and Sterzing (2022). As previously stated, the scale was designed to meet four criteria (a-d) for the illusion of understanding[2]. The theoretical derivation of the items from the definition of the illusion of understanding underpins the content validity of the scale. A more detailed discussion of the scale's validity can be found in Kulgemeyer, Hörnlein and Sterzing (2022). The scale was used after watching the video and consisted of fourteen items. As there is no longer any reference to them, the items for (b) highly comprehensible explanation must be omitted after the task. At the follow-up stage, it is not possible to make any statement about the instructional material (1a and 1b).

---

[2] The illusion of understanding is the mistaken belief that '(1) an instructional material is (a) scientifically correct and (b) a high-quality explanation in terms of comprehensibility, and (2) the learners themselves (c) understood the concept, and (d) did not require further instruction' (Kulgemeyer, Hörnlein & Sterzing, 2022, p.4).

The items include statements such as: 'I need to learn more on the concept of energy.' or 'I understood, what energy means in physics.' (according to Kulgemeyer, Hörnlein and Sterzing (2022)) The assessment of these statements was conducted on a 5-point Likert scale (ranging from 1, indicating 'I totally disagree' to 5, indicating 'I totally agree').

The reliability of the scales in study 1 ($\alpha_{video} = 0.78$; $\alpha_{task} = 0.77$; $\alpha_{follow-up} = 0.81$) and study 2 ($\alpha_{video} = 0.81$; $\alpha_{task} = 0.68$; $\alpha_{follow-up} = 0.81$) can be considered satisfactory. An exploratory factor analysis did not indicate the need to divide the scale into subscales. The scale is included in the appendix of the present paper.

*Conceptual knowledge*

The scale used to measure conceptual knowledge is derived from selected items of the 'Energy Concept Assessment' by Viering et al. (2017). This well-validated test originally consisted of 29 items, each offered various difficulty levels. The test instrument used in this study was adapted from that test. A sample item is presented in figure 3. The final scale comprises 20 single-choice items, with five items corresponding to each of the four subdomains of the energy concept, namely energy conversion, energy transport, energy dissipation, and energy conservation. The items were selected to vary in difficulty and content areas while ensuring satisfactory discriminative ability. The reliability of the scale was deemed satisfactory in study 1 (pre-test ($\alpha = 0.52$), post-test ($\alpha = 0.68$), and follow-up ($\alpha = 0.76$)) and study 2 (pre-test ($\alpha = 0.66$), post-test ($\alpha = 0.67$), and follow-up ($\alpha = $



0.69)), however, in the pre-tests the reliability has a tendency to be too low. An exploratory

factor analysis did not indicate the need to divide the scale into subscales.

**Figure 3. Sample item for concept knowledge from Viering et al. (2017) (cited from: Neumann et al., 2012), in our study only complexity information A was presented.**

*Overview of the studies, procedure, and analysis*

Both studies followed a similar experimental design (not quasi-experimental), in

which learners were randomly assigned to groups, with one key difference: Study 1 examines

the effects of high-level learning tasks, while Study 2 compares high-level and low-level

learning tasks. Study 1 includes one experimental group (receiving the video and a high-level

learning task) and one control group. Study 2 includes two experimental groups (Group 1

receiving a low-level learning task and Group 2 receiving a high-level learning task) and one

control group. Both control groups receive just the explainer video with no further learning

tasks. The test items are drawn from a standardized assessment instrument on energy and do



not use the same or similar examples as those in the videos. Our goal was to assess

conceptual knowledge rather than retention (however, we did that in other studies). Between

the initial testing phase and the follow-up test, the concept of energy was not revisited in the

lecture. No feedback on the test results was provided, and the test results were not discussed

as part of the course. However, we cannot rule out the possibility that students discussed the

test instruments in their spare time. The design of both studies is illustrated in Figure 4 for

clarity.

Hypothesis 1a can also be tested in Study 2; however, the division of participants into

three groups results in lower statistical power compared to Study 1. Comparing the findings

from both studies enhances the generalizability of the results. Due to the higher statistical

power of Study 1, Hypothesis 1b is tested only in this study, as it requires analysis at the

individual level rather than the group level.

In both studies, the pre-test and follow-up test were administered approximately one

month before or after the intervention, with each test taking 20 minutes. During the 90-

minute intervention session, participants were randomly assigned to one of two groups (study

1) or three groups (study 2). At the start of the session all participants watched an explainer

video on the concept of energy on their individual devices. Following the video (post-test I),

participants answered questions related to their belief of understanding of the explanation of

the energy concept.

Following the completion of the post-test, the treatment groups were provided with

their respective tasks and given 30 minutes to work individually before uploading their

responses. The belief of understanding was then reassessed for the treatment groups, followed

by a re-evaluation of their conceptual knowledge. In the control group, the evaluation of

conceptual knowledge occurred immediately after the assessment of the belief of

understanding that followed the video.



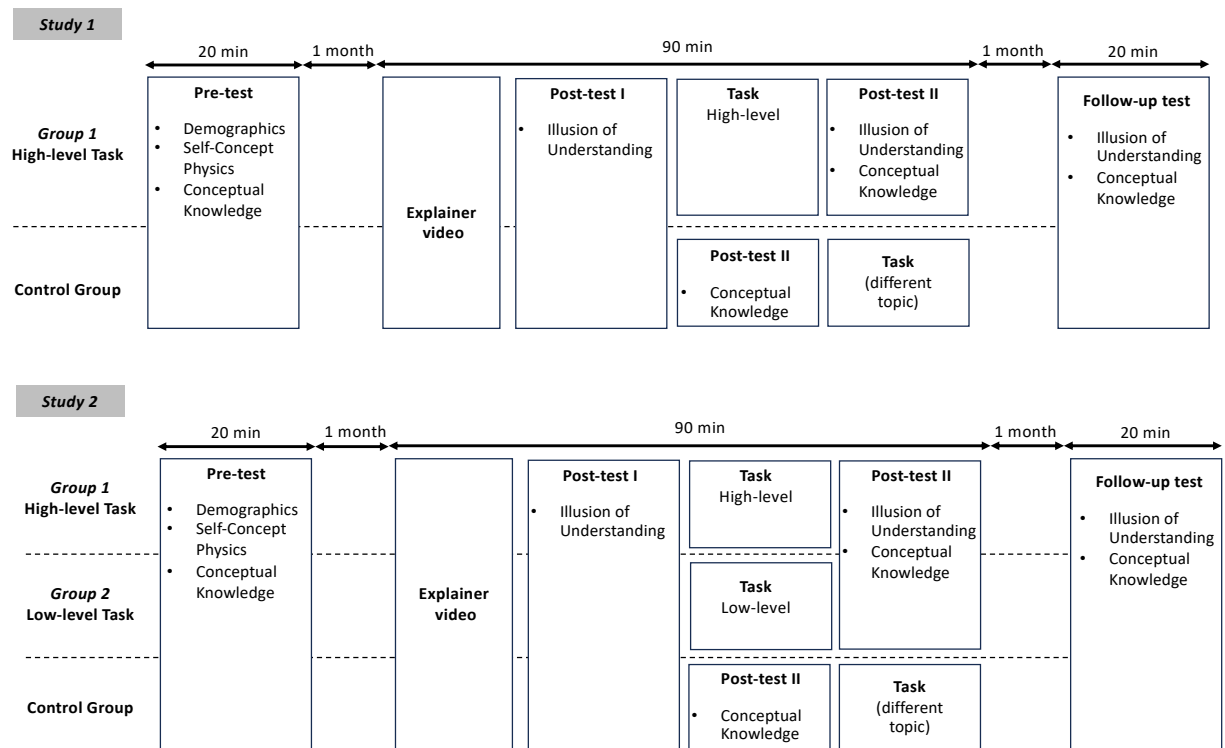

**Figure 4. Design of both studies (Study 1: above, Study 2: below).**

The instruments used to assess conceptual knowledge were consistent across all testing points and groups. However, the items related to the belief of understanding varied between assessments conducted after the video and after the task. This was due to differences in phrasing, such as 'the video was easy to understand' versus 'the task was easy to understand.' For the follow-up test, there were differences, particularly in the items related to specific material, such as video or task, as there was no longer a reference point for these.

The tests and explainer video were integrated into an online testing environment using Limesurvey. Each group accessed their respective task and could re-watch the explainer video while working on the task on a Moodle platform.

The study employed t-tests, ANOVAs, and ANCOVAs to analyze the data and evaluate conceptual knowledge and the belief of understanding. Correlations between knowledge and the belief of understanding were examined to make inferences on the illusion of understanding.



## Results of study 1

### *Descriptive statistics*

The descriptive statistics of the tested variables are presented in table 1.

**Table 1. Descriptive statistics of study variables (study 1).**

|  | High-level Group *M (SD)* (N) | Control Group *M (SD)* (N) | Overall M (*SD*) (N) | Range |
|---|---|---|---|---|
| Academic self-concept physics $\alpha = 0.91$ | 2.24 (0.83) (68) | 2.33 (0.80) (73) | 2.18 (0.79) (190) | 1-5 |
| Belief of Understanding (post I) $\alpha = 0.78$ | 3.60 (0.38) (88) | 3.59 (0.40) (96) | 3.60 (0.39) (184) | 1-5 |
| Belief of Understanding (post II) $\alpha = 0.77$ | 3.30 (0.52) (88) | -- | 3.30 (0.52) (88) | 1-5 |
| Belief of Understanding (follow-up) $\alpha = 0.81$ | 3.26 (0.62) (20) | 3.40 (0.73) (16) | 3.27 (0.66) (41) | 1-5 |
| Conceptual knowledge (pre) $\alpha = 0.52$ | 0.39 (0.15) (68) | 0.39 (0.15) (73) | 0.39 (0.15) (190) | 0-1 |
| Conceptual knowledge (post) $\alpha = 0.68$ | 0.49 (0.18) (88) | 0.50 (0.17) (96) | 0.49 (0.18) (184) | 0-1 |
| Conceptual knowledge (follow-up) $\alpha = 0.76$ | 0.51 (0.23) (20) | 0.50 (0.18) (16) | 0.51 (0.20) (41) | 0-1 |

Note: $M$ =mean; $SD$ =standard deviation; $N$ =number of participants; α =Cronbachs Alpha

### *Control variables*

The gender distribution ($\chi^2(2) = 5.05$, $p = 0.08$) and academic self-concept in physics ($t(142) = 0.53$, $p = 0.60$) did not differ significantly between the task and control groups. Additionally, no significant differences were found in conceptual knowledge in the pre-test ($t(139) = 0.14$, $p = 0.90$).

### *H1a: Combining explainer videos with high-level learning tasks reduces the illusion of understanding compared to watching the video alone.*

In the task group, scores on the belief of understanding scale before and after the task showed a statistically significant decrease with a medium effect size in a paired-samples t-test ($t(88) = 6.50$, $p < 0.001$, $d = 0.69$). This indicates that working on the task reduced the belief of understanding. The belief of understanding in the control group after watching the video did not differ from the task group before working on the task ($t(182) = -0.17$, $p =$



0.863). However, after completing the task, the task group's belief of understanding was significantly lower than that of the control group, with a medium effect size ($t(182) = 4.39$, $p < 0.001$, $d = 0.65$). Conceptual knowledge was relatively low in both groups and across all measurement points (Table 1); therefore, a lower belief of understanding can be considered more realistic. The conceptual knowledge did not differ between the control group and the task group in the pre-test ($t(139) = 0.13$, $p = .893$, $d = 0.02$) nor after the treatment ($t(183) = 0.51$, $p = .821$, $d = 0.07$).

To determine whether a belief of understanding constitutes an illusion of understanding, the belief of understanding must be high while not reflecting learners' objective conceptual understanding. To examine this, we analysed the correlation between belief of understanding and conceptual knowledge. A small but significant correlation was found between belief of understanding and conceptual knowledge immediately after watching the video in the control group ($r = 0.21$, $p = 0.045$). After working on the tasks, a stronger and significant correlation was observed in the task group ($r = 0.36$, $p < 0.001$), suggesting a more accurate alignment between perceived and actual understanding. Thus, because the task group reported a lower belief of understanding that was more closely aligned with their actual knowledge, the illusion of understanding was lower in the task group than in the control group.

### H1b: Learners with low knowledge experience a higher illusion of understanding than learners with high knowledge.

To further explore individual differences in the alignment between knowledge and belief of understanding, we conducted a quartile-based analysis, precisely replicating the approach of Dunning et al. (2003). Following their method, we divided participants into quartiles based on their knowledge performance. Both belief of understanding scores and



knowledge scores were normalized to 1, with 1 representing perfect knowledge and the maximum belief of understanding.

A clear pattern emerged in the graphical analysis (see Fig. 5, upper panel): Participants in the lower knowledge quartiles scored higher in their belief of understanding than their actual understanding warranted (indicating an illusion of understanding), whereas this discrepancy decreased in the higher quartiles, which even showed a slight tendency to underestimate their abilities (suggesting a more accurate belief of understanding). Additionally, as illustrated in the lower panel of Fig. 5, high-level tasks appeared to positively influence self-assessment, suggesting that engaging with these tasks helped participants better calibrate their beliefs about their understanding and avoid an illusion of understanding.

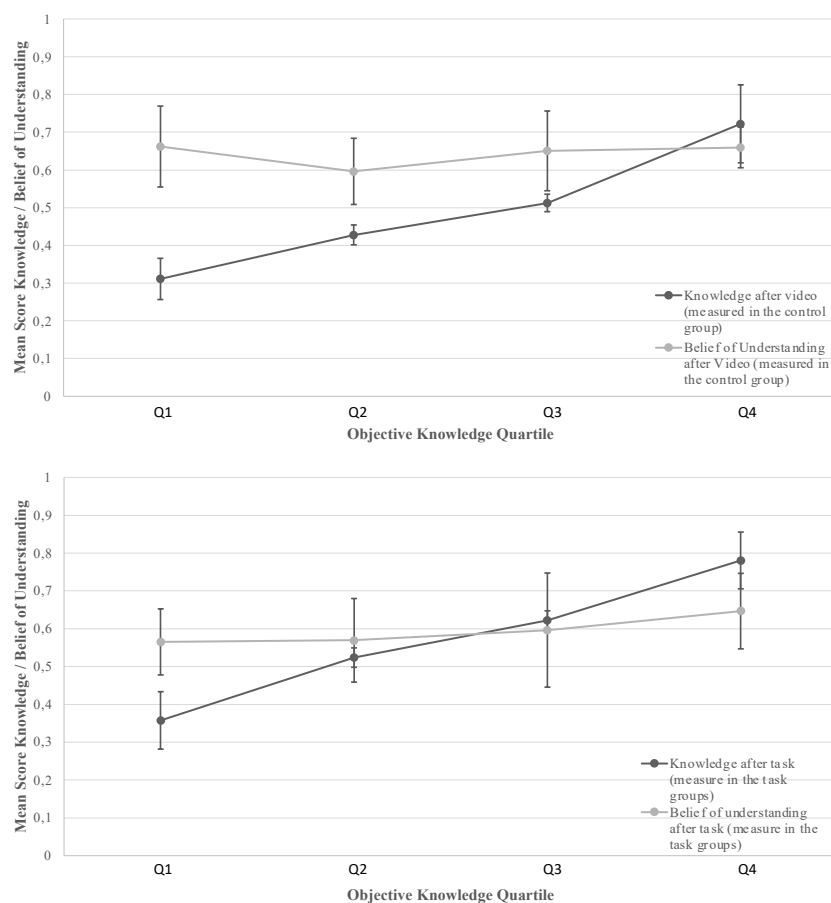

Figure 5. Mean values for objective knowledge and belief of understanding by objective knowledge quartile for study 2. The upper panel depicts the relationship after the video, while the lower panel illustrates the relationship after the task. Objective knowledge and belief of understanding are shown as mean values normalized for comparability across scales.



This influence is partially reflected in the correlational analysis. In both groups, no significant correlation was found between belief of understanding and content knowledge in the two lower quartiles (task group: $r = 0.082$, $p = 0.597$; control group: $r = -0.130$, $p = 0.373$). However, in the upper two quartiles, the correlations were stronger (task group: $r = 0.291$, $p = 0.055$; control group: $r = 0.272$, $p = 0.064$). Notably, the highest correlations were observed in the top quartile of each group (table 2).

**Table 2. Correlations between belief of understanding and content knowledge across knowledge-performance quartiles**

|  | Task group | Control group |
|---|---|---|
| Quartile 1 | $r = 0.105$, $p = 0.699$ | $r = -0.288$, $p = 0.173$ |
| Quartile 2 | $r = -0.099$, $p = 0.615$ | $r = -0.072$, $p = 0.732$ |
| Quartile 3 | $r = 0.228$, $p = 0.307$ | $r = 0.381$, $p = 0.055$ |
| Quartile 4 | $r = 0.409$, $p = 0.059$ | $r = 0.468$, $p = 0.032$ |

Note: $r$ =Pearson correlation coefficient; $p$: test for statistical significance

## Results of Study 2

### Descriptive statistics

The descriptive statistics of the tested variables are presented in table 3.

**Table 3. Descriptive statistics of study variables (study 2).**

|  | High-level Group M (SD) (N) | Low-level Group M (SD) (N) | Control Group M (SD) (N) | Overall M (SD) (N) | Range |
|---|---|---|---|---|---|
| Academic self-concept physics $\alpha = 0.89$ | 2.03 (0.58) (30) | 2.30 (0.68) (33) | 1.93 (0.63) (34) | 2.03 (0,71) (164) | 1-5 |
| Belief of Understanding (post I) $\alpha = 0.81$ | 3.57 (0.42) (30) | 3.70 (0.43) (36) | 3.57 (0.37) (35) | 3.62 (0.41) (101) | 1-5 |
| Belief of Understanding (post II) $\alpha = 0.68$ | 3.21 (0.52) (30) | 3.53 (0.46) (36) | -- | 3.38 (0.51) (66) | 1-5 |
| Belief of Understanding (follow-up) $\alpha = 0.81$ | 3.10 (0.69) (14) | 2.92 (0.51) (17) | 2.49 (0.42) (17) | 2.80 (0.63) (56) | 1-5 |
| Conceptual knowledge (pre) $\alpha = 0.66$ | 0.41 (0.18) (30) | 0.44 (0.13) (33) | 0.29 (0.12) (34) | 0.38 (0.17) (164) | 0-1 |
| Conceptual knowledge (post) $\alpha = 0.67$ | 0.57 (0.16) (30) | 0.60 (0.17) (36) | 0.49 (0.17) (35) | 0.55 (0.17) (101) | 0-1 |
| Conceptual knowledge (follow-up) $\alpha = 0.69$ | 0.51 (0.18) (14) | 0.61 (0.16) (17) | 0.42 (0.14) (17) | 0.52 (0.17) (56) | 0-1 |

Note: $M$ =mean; $SD$ =standard deviation; $N$ =number of participants; $\alpha$ =Cronbachs Alpha



*Control variables*

In study 2 as well, neither gender distribution ($\chi^2(4) = 5.04$, $p = 0.28$) nor academic self-concept in physics ($F(2, 94) = 3.029$, $p = 0.053$) differed significantly between the two experimental groups (high-level and low-level tasks) and the control group. However, the control group got outperformed by both experimental groups in prior knowledge ($F(2, 94) = 10.570$, $p < 0.001$, $\eta^2 = 0.184$). No significant difference in prior knowledge was found between the two experimental groups ($t(61) = 0.97$, $p = 0.34$).

### H2a: High-level tasks lead to a greater reduction in the illusion of understanding than low-level tasks.

To analyze the effects of high-level versus low-level tasks, a mixed ANOVA of the belief of understanding was conducted for the two experimental groups at two time points: immediately after the video and after completing the task. The analysis revealed a significant interaction effect ($F(1.00, 64.00) = 4.272$, $p = 0.043$, $\eta_p^2 = 0.063$). A more detailed analysis showed that the belief of understanding in the control group did not differ significantly from either experimental group *before* working on the task (low-level: $t(69) = -1.455$; $p = 0.150$; high-level: $t(63) = 0.018$; $p = 0.986$). However, *after* working on the task, only the high-level group exhibited a significantly lower belief of understanding compared to the control group, with a large effect size ($t(69) = 0.420$; $p = 0.676$; high-level: $t(63) = 3.237$; $p < 0.001$; $d = 0.805$). Content knowledge scores were relatively low in all groups and measurement points (Table 3); therefore, in this study as well, a lower belief of understanding can be considered more realistic. This suggests that high-level learning tasks exert a more beneficial influence on reducing an illusion of understanding.

To compare the development of the experimental groups before and after the tasks, an ANCOVA was conducted, controlling for initial belief of understanding. The results indicated a significant difference between the low-level and high-level task groups after



completing the tasks ($F(1.63) = 5.55$, $p = 0.02$, $\eta^2 = 0.08$). The difference between the

two groups was only statistically meaningful after completing the task ($t(64) = 2.61$, $p =$

$0.01$, $d = 0.65$) (figure 6). This, again, indicates that high-level learning tasks have a stronger

positive effect on reducing an illusion of understanding.

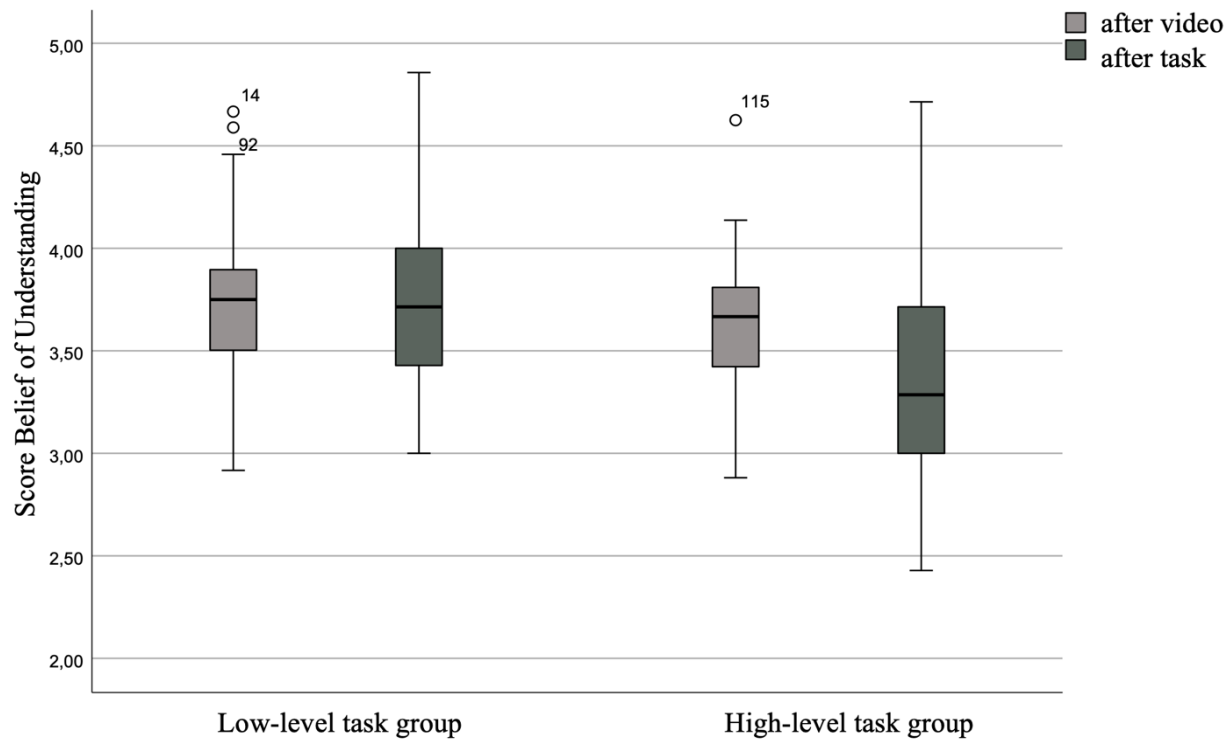

**Figure 6. Illustrating the ANCOVA: Comparison of the belief of understanding scale for the low-level task group (left) and high-level task group (right) between after the video (before the task) and after the task.**

However, no significant correlation was found between belief of understanding and

conceptual knowledge after the task in either the low-level ($r = 0.17$, $p = 0.31$) or high-

level ($r = 0.26$; $p = 0.17$) groups. This suggests that, for both groups, the belief of

understanding was still not accurate after completing the tasks.

***H2b: The reduction in the illusion of understanding over the long term persists more strongly for both types of learning tasks compared to watching just a video.***

In the follow-up test, there was no significant difference in the belief of understanding

between the low-level and high-level task groups ($t(30) = -0.51$, $p = 0.61$). Consequently,

both groups were combined and compared to the control group. In the post-test, the combined



task group exhibited a significantly higher belief of understanding than the control group, with a large effect size ($t(44) = -3.00, p = 0.004, d = 0.928$) – however, their conceptual knowledge was also significantly higher ($t(44) = -2.90, p = 0.005, d = 0.829$).

A correlation analysis for the combined task group revealed a highly significant correlation between the belief of understanding at follow-up and conceptual knowledge in the follow-up test ($r = 0.50, p = 0.005$). However, no such correlation was found in the control group at follow-up ($r = 0.03, p = 0.28$). This suggests that learning tasks have a positive long-term effect on reducing an illusion of understanding.

To explore this relationship further, we compared the correlation immediately after the task and at follow-up within the combined task group. No significant correlation was observed immediately after the task ($r = 0.24, p = 0.21$), whereas a significant correlation emerged at follow-up ($r = 0.50, p = 0.005$).

**Discussion**

The descriptive statistics indicate that immediately after watching the video, participants in both studies believed they had understood the topic and did not require further instruction. In study 1, this perception changed after completing the high-level learning task. Since the neutral score on the belief of understanding scale is 3, and all groups had a median above this threshold in the first post-test, their initial confidence suggests a possible illusion of understanding. Given the complexity of the concept of energy, it is unlikely that this high belief of understanding was justified after watching a short explainer video. Also, their post-test results in conceptual knowledge were mediocre at best. This finding aligns with previous research demonstrating the occurrence of an illusion of understanding following explainer videos (Wittwer & Renkl, 2008; Kulgemeyer & Wittwer, 2023; Kulgemeyer, Hörnlein & Sterzing, 2022).



We argue that Hypothesis 1a should be accepted: Combining explainer videos with high-level learning tasks reduces the illusion of understanding compared to watching the video alone. After completing the high-level learning tasks in Study 1, learners became more critical of their own understanding, leading to a significant decrease in their belief of understanding. While their initial belief of understanding did not differ from the control group before watching the video, it was significantly lower afterward, with a medium effect size.

However, determining whether this adjustment leads to a more accurate self-assessment requires examining the relationship between belief of understanding and conceptual knowledge. The observed correlations between these variables were small in both the control and high-level task groups, though slightly higher in the latter. This suggests that while both groups still experienced an illusion of understanding after the video and task, the learning task fostered a more realistic self-assessment. The decrease in belief of understanding indicates a greater awareness of knowledge limitations, and the higher (though small) correlation suggests improved accuracy in self-judgment.

Hypothesis 1b should be accepted tentatively: Learners with lower knowledge experience a greater illusion of understanding than those with higher knowledge. Our findings replicate key patterns from Dunning et al. (2003): participants with less knowledge tended to overestimate their abilities. Figure 5 supports this trend, as do the correlation analyses, which show stronger associations between knowledge and belief of understanding in participants with higher knowledge quartiles.

Additionally, learners in the high-level task group tended to assess their own knowledge more accurately than those in the control group. However, since most correlations were not statistically significant—likely due to limited statistical power—we accept Hypothesis 1b only as a tendency, supported by both the graphical pattern and



alignment with prior findings by Dunning et al. (2003). Our study suggests that the Dunning-Kruger effect may be a potential mechanism influencing the occurrence of the illusion of understanding. Since the mechanisms underlying the occurrence of an illusion of understanding are not yet fully understood, this is an interesting finding. Future studies with larger samples should investigate this in more detail.

For Study 2, correlational analyses should primarily compare the high-level and low-level learning task groups, as the control group had significantly higher prior knowledge. According to Hypothesis 1b, these participants should be more capable of accurately judging their understanding of the explainer video. This is a possible explanation why, in the follow-up (Hypothesis 2b analysis), the control group was significantly more critical of their own understanding than the combined task group.

Hypothesis 2a should be accepted: High-level tasks reduce the illusion of understanding more effectively than low-level tasks. Both experimental groups became more critical of their understanding after the learning task, but the high-level tasks had a stronger effect, as indicated by the ANCOVA results. This aligns with prior research suggesting that high-level tasks are more effective in improving conceptual understanding than low-level tasks (Förtsch et al., 2018). However, it adds the important perspective that not only conceptual understanding improves, but the belief of understanding also becomes more realistic.

The superior effect of high-level tasks might stem from their requirement for abstract thinking and deeper engagement, making them more effective in counteracting the illusion of understanding. However, the correlation analyses revealed no significant relationship between belief of understanding and actual knowledge for either task group. Given the lower statistical power in Study 2 compared to Study 1, the correlations still fall within a similar range, supporting the overall trend observed.



Hypothesis 2b should be accepted as well: The reduction in the illusion of understanding over the long term persists more strongly for both types of learning tasks compared to watching just a video. By the follow-up test, no significant difference in belief of understanding remained between the two task groups. Thus, for Hypothesis 2b, both groups were combined and compared to the control group.

Most notably, the combined task group showed a substantially higher correlation between belief of understanding and actual knowledge than the control group, where no correlation was observed. Additionally, the task group's self-assessment became more accurate over time, despite receiving no further instruction. This suggests that learners might need time to process the material from the task and form a more realistic self-assessment. This is an important finding because the belief of understanding becomes more accurate without further instruction, while content knowledge does not improve. However, the underlying mechanisms remain unclear and should be investigated in future studies.

In contrast, the control group maintained their illusion of understanding, as evidenced by the lack of correlation between belief of understanding and actual knowledge. The post-task and follow-up analyses indicate that learning tasks—particularly high-level ones—help learners recognize the limits of their comprehension, which literature suggests enhances learning effectiveness (Acuña et al., 2011; Fernbach et al., 2013).

Both Study 1 and Study 2 suggest that learning tasks following explainer videos lead to a more accurate self-assessment of knowledge. The fact that two separate studies, with different samples, reached the same conclusion supports the generalizability of these results. We regard this as a particular strength of the study: the main results have been replicated, which, in our experience, is rarely done in science education.



**Study limitations**

  This study is subject to several limitations, which must be acknowledged. First and foremost, it should be noted that both studies were conducted on the same single topic within the field of physics. Furthermore, the study did not control for all potential confounding variables that could have influenced the results, particularly the long-term effects of the illusion of understanding. Factors such as out-of-study learning experiences, and individual differences in student motivation could have affected the outcomes. As a result, it is imperative to exercise caution when attempting to generalize and interpret the results. It is similarly inadvisable to generalize the results to all learners, as we need more information on the learning skills of our participants and if these are representative participants of all learners. However, we attempted to select a sample that would be representative of learners in school by choosing university students who had recently completed school. Most of them had a low prior knowledge of physics. As the study was conducted in a laboratory setting, its external validity is relatively low. Another point that requires caution in interpreting the data is the relatively low reliability of less than 0.7 for some scales, which may, for example, affect the accuracy of dividing learners into knowledge quartiles. A significant limitation of the study is participant attrition. Due to the voluntary nature of participation, many students did not complete all stages of the study, resulting in missing data.

**Study implications**

  In conclusion, the findings of this study suggest that there may be potential risks associated with relying solely on explainer videos for student learning. After watching an explainer video, students may not accurately assess their own understanding. Actively engaging in learning tasks has the potential to counteract the illusion of understanding that videos may induce. Of course, other countermeasures are also possible—likely through



increased cognitive engagement, as suggested by Chi et al. (1994)—and learning tasks represent only one way to achieve this.

When considering post-video effects, it is important to differentiate between low-level and high-level tasks. In our study, a more realistic self-assessment of knowledge emerged over the long term, regardless of task type. Future studies should examine this effect in greater depth: Even though there was no further instruction on this topic—and, consequently, the knowledge scores did not improve from post-test to follow-up—the belief of understanding became more accurate. This pattern suggests that learners may require some latency for their self-evaluations to settle; however, this adjustment occurred only in the task group, indicating that the tasks may have helped to initiate this process.

Based on these findings, we view the incorporation of learning tasks after videos as a promising strategy; however, we refrain from recommending a specific task type. Implementing such tasks does not necessarily require a flipped-classroom approach, although such an approach is certainly possible. We would argue that low-level tasks, such as summarizing video content, may be particularly useful because they require little preparation from instructors while still engaging students in a meaningful way. The recommendation that videos should not be used in isolation remains valid—passive video consumption should be avoided when the goal is effective learning of physics content (Kulgemeyer, 2018). Our study adds an important argument for embedding videos within learning tasks that goes beyond content acquisition: we consider it essential to make learners aware of their deficiencies so that these can be addressed through targeted learning activities.

Future research should further examine the mechanisms by which post-video tasks mitigate the illusion of understanding. It remains unclear whether this effect is driven primarily by self-assessment processes, metacognitive engagement, or a combination of both.



A deeper understanding of these mechanisms may lead to more targeted strategies for improving students' learning experiences.

**Disclosure Statement**

The authors have no competing interests to declare that are relevant to the content of this article.

**Data availability**

The data that support the findings of this study are not openly available due to reasons of sensitivity and are available from the first author upon reasonable request.

**Ethical Statement**

The study was conducted in accordance with the ethical guidelines of the University XX, including obtaining informed consent. All participants participated voluntarily.